\documentclass[journal]{IEEEtran}
\usepackage{cite}
\usepackage{graphicx}
\usepackage{subfigure}
\usepackage{amssymb} 
\usepackage[cmex10]{amsmath} 
\usepackage{multirow}
\usepackage{array}
\usepackage{float}
\usepackage{perpage}
\MakeSorted{table}
\usepackage{afterpage}

\begin{document}

\title{Theoretical grounds for the propagation of uncertainties in Monte Carlo particle transport}

%
%

\author{Paolo Saracco, Maria Grazia Pia, Matej Batic 
\thanks{Manuscript received June 15, 2013.}
\thanks{P. Saracco and M. G. Pia are with INFN Sezione di Genova, Genova, 16146 Italy (telephone: +39-010-3536210, e-mail: Paolo.Saracco@ge.infn.it, Maria.Grazia.Pia@cern.ch).}%
\thanks{M. Batic was with INFN, Genova, 16146 Italy and now is with Sinergise, Ljubljana, Slovenia.}%
}

\maketitle

\begin{abstract}
We introduce a theoretical framework for the calculation of uncertainties affecting
observables produced by Monte Carlo particle transport, which derive from
uncertainties in physical parameters input into simulation.
The theoretical developments are complemented by a heuristic application, which
illustrates the method of calculation in a streamlined simulation environment.


\end{abstract}


\section{Introduction}

\IEEEPARstart{U}{ncertainty} 
quantification (UQ) is a fast growing sector in
interdisciplinary research: its applications span political science
\cite{Wal2003}, computational biology\cite{Mar2008}, climate science
\cite{Lev1999}, economic and financial processes \cite{Sus2004}, industrial and
civil engineering \cite{Ber2010}, as well as many other disciplines.
In a broad sense uncertainty quantification is a domain of applied
mathematics; the variety of its applications has promoted a large number of
approaches and methods to address the problem.

Uncertainty quantification is an issue in scientific computing.
Reviews of ongoing research in this field can be found in
\cite{Obe2002,swiler_2009,helton_2011,pnnl_20914,oberkampf_book};
interested readers can find further information in the bibliography of the cited
references.
Software systems, such as DAKOTA \cite{dakota}, PSUADE \cite{psuade}
and similar codes reviewed in \cite{pnnl_20914}, have been developed to
facilitate this task, mainly focusing on methods and algorithms for sensitivity
analysis and statistical evaluation of uncertainties.

Uncertainty quantification is especially relevant to physics simulation, where
the ability to estimate the reliability of simulated results is critical to
establish it as a predictive instrument for experimental research.
Nevertheless, relatively limited attention has been invested so far into the
problem of quantifying the uncertainties of the outcome of Monte Carlo particle
transport in general terms.

Investigations of uncertainty quantification in the domain of Monte Carlo
particle transport mainly concern applications to nuclear power systems, such as
\cite{avramova_2010,sternat_2011,diez_2012,mccard_2012}.
Common experimental practice in other application areas, such as high energy
physics experiments, focuses on the validation of specific use cases by direct
comparison of simulation results and experimental measurements: representative
examples of this practice can be found in \cite{marshall_2010,benekos_2008,
banerjee_2011,rovelli_2008, easo_2005}, which concern experiments at the LHC
(Large Hadron Collider).
Hardly any effort has been invested so far in estimating the predictive
capabilities of simulation codes, such as EGS5 \cite{egs5}, EGSnrc \cite{egsnrc}, 
FLUKA \cite{fluka1,fluka2}, Geant4 \cite{g4nim,g4tns}, ITS5 \cite{its5}, 
MCNP \cite{mcnp5,mcnpx27e} or PENELOPE \cite{penelope}, commonly
used in these experiments: this ability would be useful in experimental
scenarios where direct validation of simulation use cases would be difficult or
not practically feasible, for instance in some space science projects,
astroparticle physics experiments and medical physics investigations, as well as
in the process of detector design, where the hardware that is simulated may not
yet exist. 

This paper defines a theoretical foundation for the calculation of the
uncertainties affecting simulated observables, which are a consequence of the
uncertainties affecting the input to the simulation itself.
This capability is the basis for establishing the predictive reliability of Monte Carlo 
transport codes in experimental practice.

The quantification of the uncertainties that affect the results of Monte Carlo
simulation as a consequence of the uncertainties associated with its physical
input is a vast and complex problem, which requires extensive scientific
research.
This paper is not intended to present an exhaustive solution to the problem, nor
to document applications to real-life experimental scenarios simulated with
general purpose Monte Carlo transport codes;
its scope is limited to setting a theoretical ground, which to the best of our
knowledge has never been previously documented in the literature, to enable
further conceptual and mathematical progress in this field in view of future
experimental applications.

 A preliminary report of this study is documented in \cite{nss2012_uq}.

\section{Overview of the problem domain}
\label{sec_domain}

Uncertainty quantification in the context of a computational system is the
process of identifying, characterizing, and quantifying those factors that could
affect the accuracy of the computational results \cite{oberkampf_book}.

Uncertainties can arise from many sources; the computational model propagates
them into uncertainties in the results.
This problem is usually referred to as forward uncertainty quantification.
In experimental practice one encounters also the problem of backward uncertainty
quantification, i.e. the assessment of the uncertainties that may be present in
a model: this issue is of raising interest for its applicability in robust
design, as it concerns the ability of making a rational choice among different
conceptual designs that can be drafted.

Uncertainties in the results can derive from the conceptual model
upon which a computational system is constructed, the formulation of the model
in the software and the actual computation process.
Possible sources of uncertainties are characterized in \cite{Ken2001}:
\begin{enumerate}
\item {\em parameter uncertainty} identifies situations where some of the
computer code inputs are unknown;
\item {\em model inadequacy} may derive from {\em structural uncertainty}
related, for instance, to approximations in the used model, and from {\em
algorithmic uncertainty} related to the numerical methods employed to solve the
model;
\item {\em residual variability} occurs when the process itself may be inherently
unpredictable and stochastic, or the model itself is not fully specified;
\item {\em parametric variability} concerns use cases where some of the inputs
are intentionally uncontrolled or unspecified, therefore they contribute a
further component of uncertainty to the predicted process.
\end{enumerate}

Parameter uncertainty plays a major role in Monte Carlo simulations of particle
transport, since in practice all the physical input to the simulation is
affected by uncertainties: the cross sections determining particle interactions
either derive from interpolations of experimental measurements, which are
affected by uncertainties, or from analytical models, which in turn involve
experimental or theoretical uncertainties.
In addition, parameter uncertainties are involved in the model of the
experimental set-up: they concern the geometrical dimensions and material
composition of the apparatus, and the conditions of its operation, such as
electromagnetic fields, pressure and temperature.

Algorithmic uncertainty, as defined by \cite{Ken2001}, derives from the
numerical methods employed to solve the model: in this acception it is implicit
in the Monte Carlo transport process, which is a statistical determination of
some physical quantities (densities, fluxes, energy deposition etc.) of
experimental interest through sampling methods.

Structural uncertainties may be present in the physics models on which the
transport is based: for instance, some effects may be neglected a priori in the
formulation of the models (e.g. those deriving from the molecular composition of
materials, or their solid structure), or the models may involve assumptions that
are questionable in some energy range of the transported particles.
Condensed history methods fall into this category: they are approximate Monte
Carlo methods to deal with physics processes affected by infrared divergence and
involving a large number of collisions with very small changes in direction and
in energy, which would require a prohibitive investment of computing resources
to be treated exactly.
Moreover, the concept itself of particle transport is an approximation to the real
world: Monte Carlo codes for particle transport are based on the assumption of
classical particles moving between localized interaction points, while quantum
mechanics is implicit in the determination of the cross sections involved in the
process.

Whereas these concepts provide guidance in the identification of possible
sources of uncertainties, a rigid scheme of classifications can hardly capture
the complexity of the problem domain.
For instance, parameter and structural uncertainties are often intertwined in
Monte Carlo codes for particle transport, as input data often actually embed
physics models: theoretical calculations that describe particle interactions
with matter are usually tabulated in data libraries, or transformed into look-up
tables in the initialization phase of the Monte Carlo simulation to reduce the
computational burden of the simulation.
These data implicitly contain modeling assumptions and approximations;
the associated uncertainties, although treated as parameter uncertainties,
may encompass a component of model inadequacy.

The problem domain encompasses correlated and uncorrelated uncertainties. 
In the context of Monte Carlo simulation for particle transport, correlated
uncertainties may originate from systematic errors in the physical data required
for the simulation, while uncorrelated ones may be associated with statistical
errors of input experimental parameters.
Mixed and more complex situations may also occur, for instance where
experimental data with their associated uncertainties are used in connection
to physical models, whose validity is in turn variable. 

\section{Strategy of this study}
\label{sec_strategy}

This paper deals with forward propagation of uncertainties in Monte Carlo
simulation.

\begin{figure}[!tbp]
\centering
\includegraphics[width=9cm]{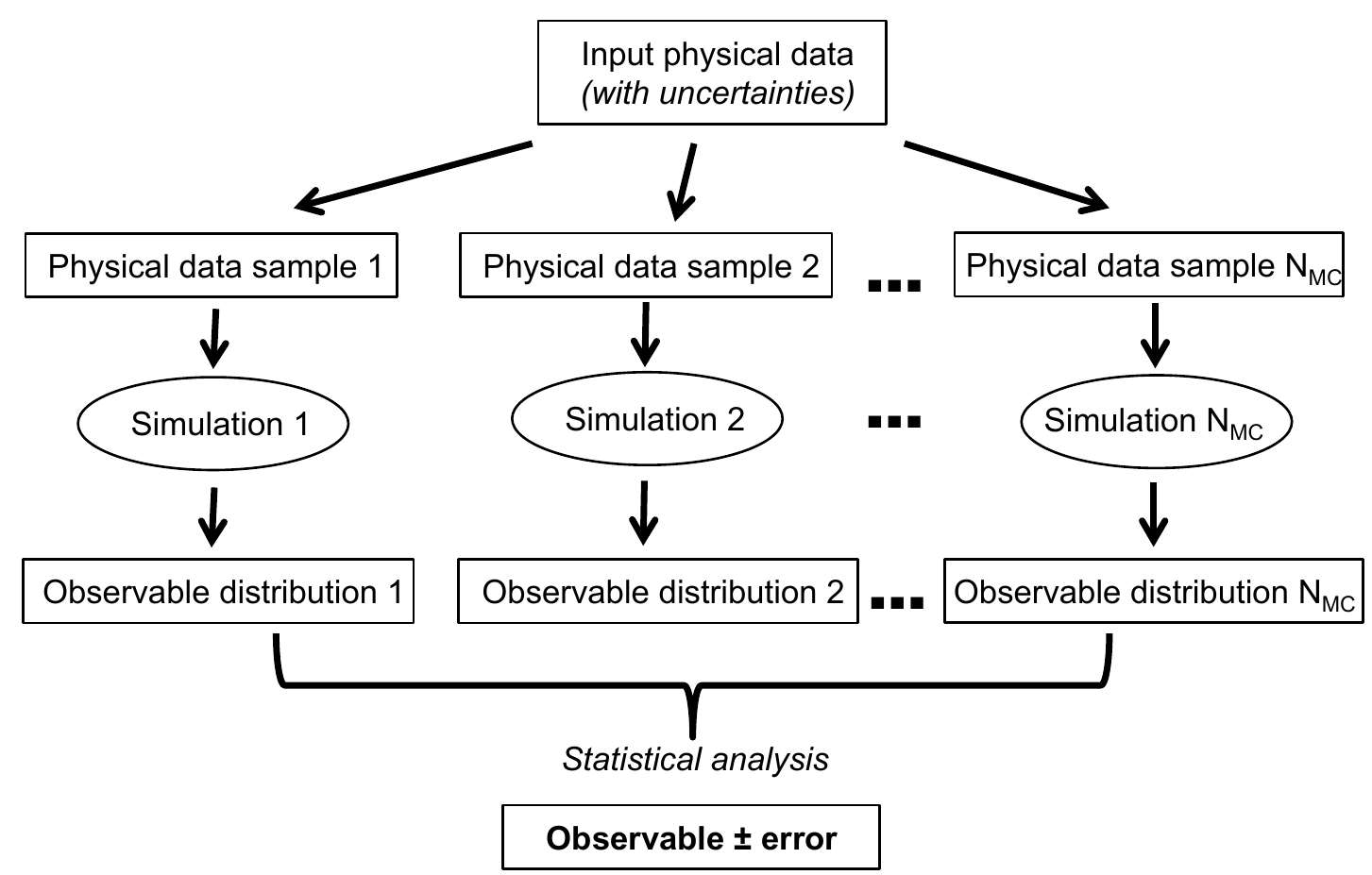}
\caption{Conceptual scheme of the uncertainty quantification process.}
\label{Fig:over}
\end{figure}

The conceptual abstractions involved in the process of uncertainty propagation
and their functional relationships are illustrated in Fig.~\ref{Fig:over}.
Input parameters of the simulation are affected by uncertainties: conceptually,
this situation corresponds to executing simulations with many possible input
parameters, which vary compatible with their associated probability
distributions, and produce many possible outputs.
The statistical properties of the outcome of the simulations represent the
effect of input uncertainties.

The following section elaborates a mathematical foundation for the calculation
of uncertainties in observables produced by a Monte Carlo simulation, which
derive from the uncertainties affecting quantities fed into the particle
transport process.
The adopted approach aims at a complete determination of the probability
distribution function (PDF) of the output observable;
in this respect, it differs from other approaches to uncertainty quantification,
such as those pursued by DAKOTA, PSUADE and similar software tools, which focus
on the management of large ensembles of calculations used in a statistical
evaluation of uncertainty.

Although the theoretical foundations and calculation methods developed in this
paper are generally applicable to any input parameters, the following
discussions are mainly focused on the quantification of the effects of physical
uncertainties, and their interplay with the algorithmic uncertainties associated
with the Monte Carlo sampling process.

Theoretical investigations are carried out in a simplified calculation
environment, which retains the essential conceptual features characterizing the
problem domain.
The scheme for uncertainty propagation is initially developed for quantifying
the effects on an observable associated with the uncertainties affecting a
single parameter.
The conceptual foundation established in the present paper is the basis for
further calculations in more complex environments, representing more realistic
experimental scenarios, which are intended to follow.

The theoretical elaboration described in section \ref{sec_theory} establishes
that it is in principle possible to disentangle the effect of parameter
uncertainties from algorithmic uncertainties: it is possible to determine a
probability distribution of the outcome of the original physical
problem deriving from parameter uncertainties alone.
By algorithmic uncertainties we mean those related only to the numerical methods
employed to solve the underlying transport equations, i.e. from Monte Carlo
sampling, as defined in section \ref{sec_domain}.
The effect of Monte Carlo sampling consists of blurring the distribution
resulting from the propagation of parameter uncertainties with some statistical
noise.

The next stage of investigation consists of devising a procedure, based on these
findings, to calculate in practice the probability distribution of the possible
outcomes of an observable within a desired margin of statistical error.
The possibility of performing this calculation by means of a small number of Monte
Carlo simulations is examined in section~\ref{sec_statcalc}, while other practical aspects
of uncertainty quantification are discussed in section~\ref{sec_methods}. 

The discussion of the theoretical aspects of the problem is supported by a
heuristic investigation, documented in section~\ref{sec_application},
which illustrates empirically the process of uncertainty propagation and the
conclusions deriving from its mathematical foundation.

In the context of this paper it is assumed that uncertainties affecting the
physical input to Monte Carlo particle transport are known.
Their quantitative knowledge is established by evaluation procedures applied to
data libraries of experimental origin, such as some components of
\cite{ENDFBVII}, and dedicated validation tests, such as those documented in
\cite{tns_rayleigh,tns_beb,tns_binding,tns_pcross,tns_pixe,tns_relax_prob,tns_relax_nist}.
This assumption does not reflect the real status of general purpose Monte Carlo
codes for particle transport: although some of them have been widely used for
decades, the validation of their physics modeling is still in progress, and
quantitative assessments of the uncertainties associated with their basic
physics parameters are scarcely documented in the literature.
This process is further hindered by the presence of epistemic uncertainties,
i.e. intrinsic knowledge gaps \cite{tns_epistemic}.
The current incomplete quantification of the physical input uncertainties in
Monte Carlo particle transport codes does not hinder the conceptual foundation
of the theoretical scheme discussed in this paper.
Nevertheless, such knowledge is required for practical applicability of the
findings of this paper, i.e. for the calculation of uncertainties affecting
output observables.

\section{Theory}
\label{sec_theory}

The subject under study is how the sources of uncertainty in the input to the
simulation, wherever they originate, are mathematically mapped to uncertainties
in the simulation results.
We show that under some general conditions, which are elucidated in the
following, Monte Carlo simulation transfers the distribution of input data
uncertainties into the distribution of the results, preserving the same
functional form or with a form derived from it, and adding some statistical fluctuation.

\subsection{Propagation of parameter uncertainties}
\label{sec_theoeq}

If one runs many Monte Carlo simulations, each one encompassing $N$ events,
varying one of the physical parameters involved in particle transport
(identified here with $\Sigma$) in some interval with probability distribution
$f\left(\Sigma\right)$, then the final distribution $G(x)$ for the value $x$ of any desired
observable $X$, which depends on the value of $\Sigma$, is expected to
be
\begin{eqnarray}
G(x)\simeq\qquad\qquad\qquad\qquad\qquad\qquad\qquad\qquad\qquad\quad\label{eqn:start}\\
\int_{-\infty}^{+\infty}\,d\Sigma\,f\left(\Sigma\right)
\exp\left[-\frac{\displaystyle \left(x-x_0(\Sigma)\right)^2}{\displaystyle 2\sigma_{x_0}^2/N}\right]
\sqrt{\frac{\displaystyle N}{\displaystyle 2\pi\sigma_{x_0}^2}}
\nonumber
\end{eqnarray}
where we finally integrated over the possible values of the physical parameter
$\Sigma$ with their probabilities. Here $x$ is a stochastic variable because it represents
the sampled statistical mean of the contributions to the requested observable from the $N$ events and
$\sigma_{x_0}^2$ is its sampled variance.
This result derives from the Central Limit Theorem, if $N$ is sufficiently large
to make $\sigma_{x_0}^2$ independent from $N$ itself.

When writing this equation, we made a further assumption:
namely that a function $x_0\left(\Sigma\right)$ exists, which relates any
input value of $\Sigma$ to a corresponding value for the peak position of the
observable means distribution.

This assumption relies on the interpretation of the process of simulation
as a surrogate for the solution of the Boltzmann transport equation.
The function $x_0\left(\Sigma\right)$ represents the parametric dependence of the physical
solution we produce through the simulation to a variation of the input
parameter.
Full knowledge of $x_0\left(\Sigma\right)$ is equivalent to the ability of
solving the transport equation.

If this function is invertible - a fact that, as we shall see later in Sect.\,\ref{sec_appl_linear},
can be established when necessary - we may change the variable of integration
from $\Sigma$ to $x_0$; using $\Sigma=\Sigma(x_0)$, 
we obtain in the limit $N\to\infty$
\begin{eqnarray}
G(x)&=&\int_{-\infty}^{+\infty}\, d\Sigma\,f(\Sigma)\delta(x-x_0(\Sigma))\label{eqn:fin}
\\ &=& \nonumber
\left|\frac{d\Sigma(x)}{dx}\right|f\left(\Sigma(x)\right)
\end{eqnarray}
where we used the identity
\begin{eqnarray}
f(x)=\lim_{\sigma\to 0}\sqrt{\frac{\displaystyle 1}{\displaystyle 2\pi\sigma^2}}\int_{-\infty}^{+\infty}\,dx_0\,f\left(x_0\right)
\exp\left[-\frac{\displaystyle \left(x-x_0\right)^2}{\displaystyle 2\sigma^2}\right]\nonumber
\end{eqnarray}
stemming from the exponential representation of Dirac $\delta$ function. 
If the function $x_0(\Sigma)$ is invertible only over subintervals of the
variability interval for $\Sigma$ - that is the same physical solution $x_0$
results from different values of the input parameter $\Sigma$ - we must slightly
modify (\ref{eqn:fin}) by summing over all the possible solutions of the
equation
$\Sigma=\Sigma(x_0)$.

Equation (\ref{eqn:fin}) states that we are able to exactly know how the input
probability distribution of physical data transfers into the final result,
provided we are able to determine the unknown function $\Sigma(x)$.
This conclusion is the theoretical foundation 
for uncertainty quantification. 

We derived equation (\ref{eqn:fin}) in the context of Monte Carlo simulation;
nevertheless, through the limit $N\to\infty$ the formula has dropped any direct
reference to the Monte Carlo simulation environment and expresses a general
property of forward propagation of uncertainty for an arbitrary deterministic
problem.
In other environments this expression is known as the Markov formula\cite{GUM,Cox}.

The theory here presented is valid independently from the number of unknowns:
the first line of (\ref{eqn:fin}) holds even if
$\Sigma\,\longrightarrow\,\{\Sigma_1,\ldots,\Sigma_M\}$ and
$x\,\longrightarrow\,\{x_1,\ldots,x_K\}$. 

\subsection{Verification}
\label{sect:verify}
The correctness of the method outlined in section \ref{sec_theoeq} can be
verified in a scenario where the outcome of a Monte Carlo calculation is an
exactly solvable problem.

As an example, we consider the calculation of the area of a circle; we assume
that the radius of the circle is known with some uncertainty.
The solution $A(R)=\pi R^2$ is invertible as $R(A)=\sqrt{A/\pi}$ for $R>0$ (the only physically acceptable solution), where $A$ is
the area and $R$ is the radius.

On the basis of equation (\ref{eqn:fin}), the expected probability distribution
function for the area of a circle with input uncertainty on the measure of
its radius is:
\begin{equation}
G(A)=\frac{dR(A)}{dA}f(R(A))
\end{equation}
where $f(R)$ is the PDF for the unknown input radius. 
If, for example, $f(R)$ is a flat distribution with 
$R_{\rm min}\le R\le R_{\rm max}$, then
\begin{equation}
G(A)=\frac{1}{2\pi}\sqrt\frac{\pi}{A}\theta(A-A_{\rm min})\theta(A_{\rm max}-A)
\label{eqn:theopre}
\end{equation}

\begin{figure}[!tbp]
\centering
\includegraphics[width=9cm]{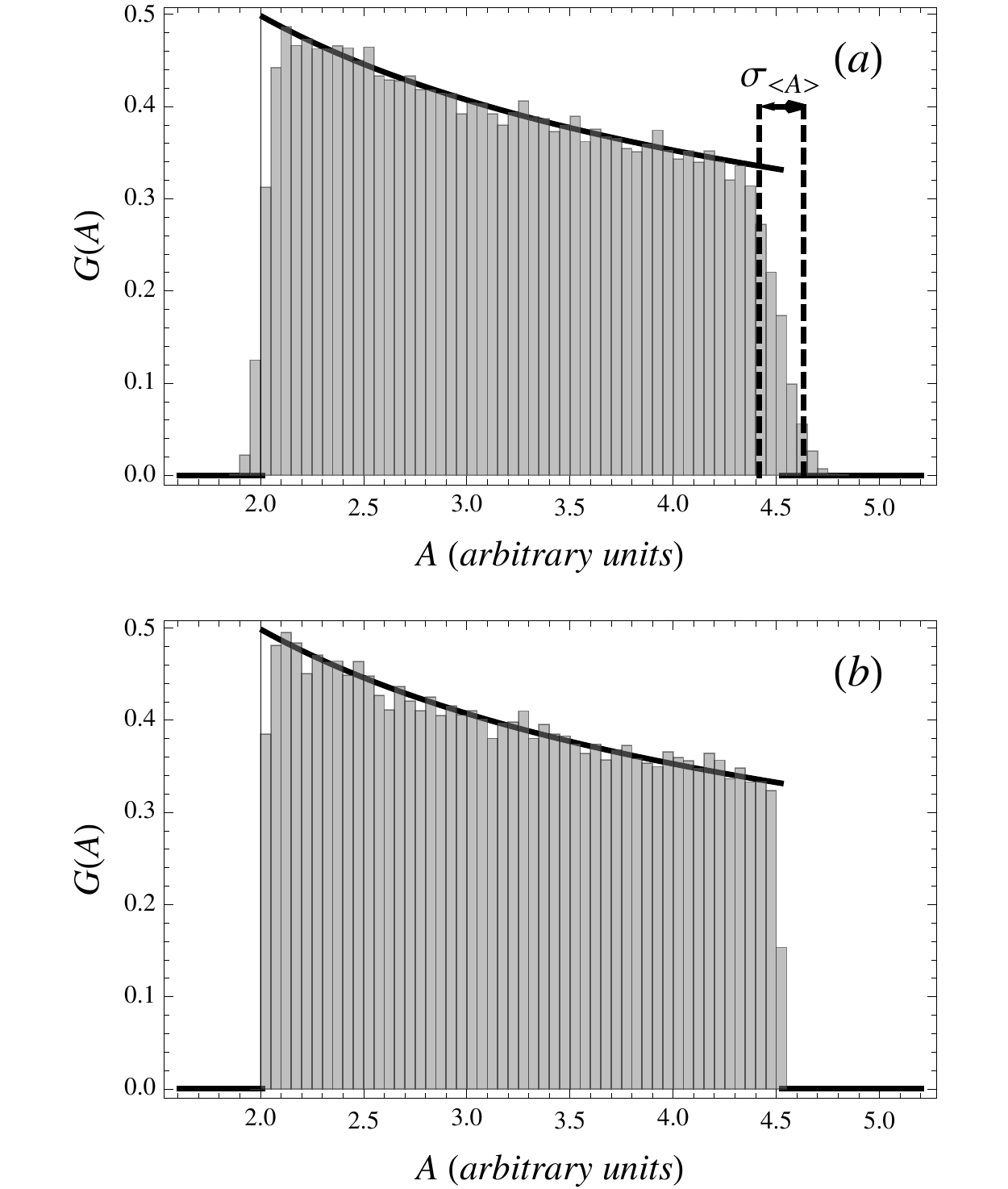}
\caption{ 
The expected PDF $G(A)$ for the area of a circle as a function of $A$
with randomly chosen radii $0.8\le R\le 1.2$ (thick black curve), together with the
empirical distributions (grey) coming from 100000 Monte Carlo estimates of the same
area.
In (a) the Monte Carlo simulation encompasses 500 samples, in (b) 50000.
In (a) the standard deviation on
$<A>$ at the fixed value $A=A_{\rm max}$ is also shown with two dashed lines}
\label{Fig:Metropolis}
\end{figure}

We consider a test case where the radius is uniformly distributed in the
interval $0.8\le R\le 1.2$ (in arbitrary units), so that $A_{\rm min}=2.011$ and
$A_{\rm max}=4.524$ (in arbitrary units).
The results of the Monte Carlo calculation of the area are compared with the
theoretical prediction of equation (\ref{eqn:theopre}) of its PDF in
Fig.~\ref{Fig:Metropolis}: the two histograms correspond to the empirical
distribution of the area resulting from 500 and 50000 Monte Carlo samples
respectively, each sample encompassing 100000 estimates, while the thick
curve represents the theoretical PDF as in (\ref{eqn:theopre}).
The consistency of the PDF of the simulated observable with the theoretical
prediction ($A^{-1/2}$) is qualitatively visible, and appears more evident in the plot
resulting from the larger Monte Carlo sample; it is quantitatively confirmed
with 0.05 significance by a statistical comparison of the results: the p-values
are respectively 0.083 and 0.910 for a Cramer-Von-Mises test with a bin width of
0.01.
The blurring effect deriving from the process of Monte Carlo calculation
is also manifest.

This verification test is also instrumental to highlight an issue, which will be
discussed more extensively in section \ref{sec_application}: the interplay of
different statistical errors in the process of Monte Carlo sampling.

The upper plot of Fig.~\ref{Fig:Metropolis} also shows the standard deviation on
$<A>$ at the fixed value $A=A_{\rm max}$ with two dashed lines.
This should help the reader to understand that within the process of Monte Carlo
sampling the exact result expressed in equation (\ref{eqn:fin}) implies the
mixing of two different statistical errors: one deriving from the finiteness of
the number of events $N$ in each MC simulation (see equation
\,(\ref{eqn:start}), and one stemming from the finiteness of the number of Monte
Carlo really run.
The effect of the former is clearly visible on the sides of the empirical
distribution, which are exponentially smeared around the exact values $A_{\rm
min}$ and $A_{\rm max}$ by an amount $\sigma_{<A>}=\sigma_A/\sqrt N$.
The comparison of Fig.~\ref{Fig:Metropolis}(a) and \ref{Fig:Metropolis}(b) makes
also visually evident the limiting process in going from (\ref{eqn:start}) to
(\ref{eqn:fin}): in (b) $\sigma_{<A>}$ is reduced by a factor 10 with (therefore it
cannot be visually displayed in the plot), because we used 100 times more Monte 
Carlo samples.
The oscillations of the empirical distribution around its theoretical value
within the interval of allowed values $A_{\rm min}<A<A_{\rm max}$ are, instead, a
consequence of the finiteness of the number $N_{\rm MC}$ of the number of Monte Carlo
simulations really run: their amplitude is ruled by $1/N_{\rm MC}^{1/2}$.

The theoretical elaboration in section~\ref{sec_theoeq} has reduced the problem
of uncertainty quantification from the computationally demanding production of
the empirical distribution, requiring a large number of Monte Carlo simulation runs
(as depicted in Fig.~\ref{Fig:over}), to the determination of the parameters
fixing the form of the expected theoretical distribution: in this verification
test they are $A_{\rm min}$, $A_{\rm max}$ and the exponent of its power form.
In this test case they are all known, since the given problem is analytically
solvable; in a realistic situation these parameters must be determined through
the simulation itself.
In section~\ref{sec_appl_pdf} we will show that it is possible to determine
these parameters with an accuracy ruled by $\sigma_{<x_0>}=\sigma_{x_0}/\sqrt N$
through a small number of Monte Carlo runs. in this example  the observable $x_0$ coincides
with the area $A$.

\section{Statistical estimation}
\label{sec_statcalc}

Equation (\ref{eqn:fin}) is fundamental to the purpose of uncertainty
quantification, as it relates the functional form of the output PDF $G(x)$ to
those of $f(\Sigma)$, which is a necessary input, and of $\Sigma(x_0)$ or
$x_0(\Sigma)$.
Nevertheless, equation (\ref{eqn:fin}) is of limited practical usefulness to
quantify the uncertainties of observables resulting from Monte Carlo simulation,
since it is equivalent to solving the transport equation: if one were able to do
this, one would not need any Monte Carlo simulation at all.

A remark is here in order: equation (\ref{eqn:fin}) is an exact relation, even if we
derived it starting from (\ref{eqn:start}) which is valid only for a MC simulation; 
knowledge of $x_0(\Sigma)$ enables to know the required form of $G(x)$ from which
any statistical information about the observable $x$ can be extracted. Then in principle whichever solver
for the transport equation can be used to this purpose, not only MC simulation. 
Most part of the discussion in this Section and in Sect.\,\ref{sec_methods} can be translated accordingly, by substituting
"statistical accuracy" with "numerical accuracy" and "MC simulation" with "numerical solver for the transport equation".

A practical approach to address the problem consists of using the Monte Carlo
simulation process itself to statistically estimate the unknown function
$\Sigma(x_0)$:
this is possible, because any single Monte Carlo run, executed with a fixed
value of $\Sigma$, gives a statistical estimate of the value of $x_0(\Sigma)$.
Uncertainty quantification requires the ability to perform this estimate within
some predefined margin of error.
It is worth noting that the task discussed here is not the prediction of the
simulated PDF for $x$ resulting from a large number of Monte Carlo simulations,
which would be computationally expensive, but only the calculation of the
parameters determining the final exact PDF $G(x)$ defined in
(\ref{eqn:fin}).

This task becomes particularly simple when the relation $x_0(\Sigma)$ is linear,
or can be treated as approximately linear, as it happens for instance in the example we give in the following (see Fig.~\ref{fig_lin}).
In this case equation~(\ref{eqn:fin}) reduces to
\begin{equation}
G(x)\propto f\left(\Sigma(x)\right)\,,
\label{eqn:finsimp}
\end{equation}
which represents the PDF for the observable means as a linear map of the
PDF for the unknown input physical parameter.
The two parameters characterizing the linear map can be practically determined with
any predefined margin of statistical accuracy by running two Monte Carlo
simulations for two distinct values of the input variable $\Sigma$.

While it is unlikely that $x_0(\Sigma)$ is linear in general, any function can
be approximated by a linear relation in a sufficiently small interval \cite{LuSte}.
For the practical purpose of reducing equation~(\ref{eqn:fin}) to the easily
manageable form of equation~(\ref{eqn:finsimp}), it is sufficient that
$x_0(\Sigma)$ can be considered linear over an interval relevant to the
simulation scenario: over the whole variability interval, if the input
distribution is supported on a bound interval, otherwise over
some predefined confidence interval.

The hypothesis of a linear approximation of the functional form of $\Sigma(x)$
in the experimental scenario under study can be verified by running Monte Carlo
simulations for a few suitably chosen points in the variability interval for
$\Sigma$.
As a minimal requirement for estimating the suitability of a linear
approximation one could envisage three simulations, for instance at the two
interval extrema and at the mode, while a more powerful test of the hypothesis
of linearity could be performed with a larger number of simulation runs.

The same conceptual procedure applies if, instead of a linear relationship, one
considers other computationally manageable approximations, for instance a power
representation of the form
\begin{equation*}
x_0(\Sigma)=\sum_{k=0}^m a_k(\Sigma-<\Sigma>)^k\,.
\end{equation*}

The procedure described above to statistically determine the parameters of the
theoretical probability distribution function of an observable is
computationally less expensive than determining the probability distribution
function of the observable from a large number of Monte Carlo simulations, each
one performed with a different value of the input parameter consistent with its
uncertainty.

\section{Application of the method}
\label{sec_application}

\begin{figure}[!bp]
\centering
\includegraphics[width=9cm]{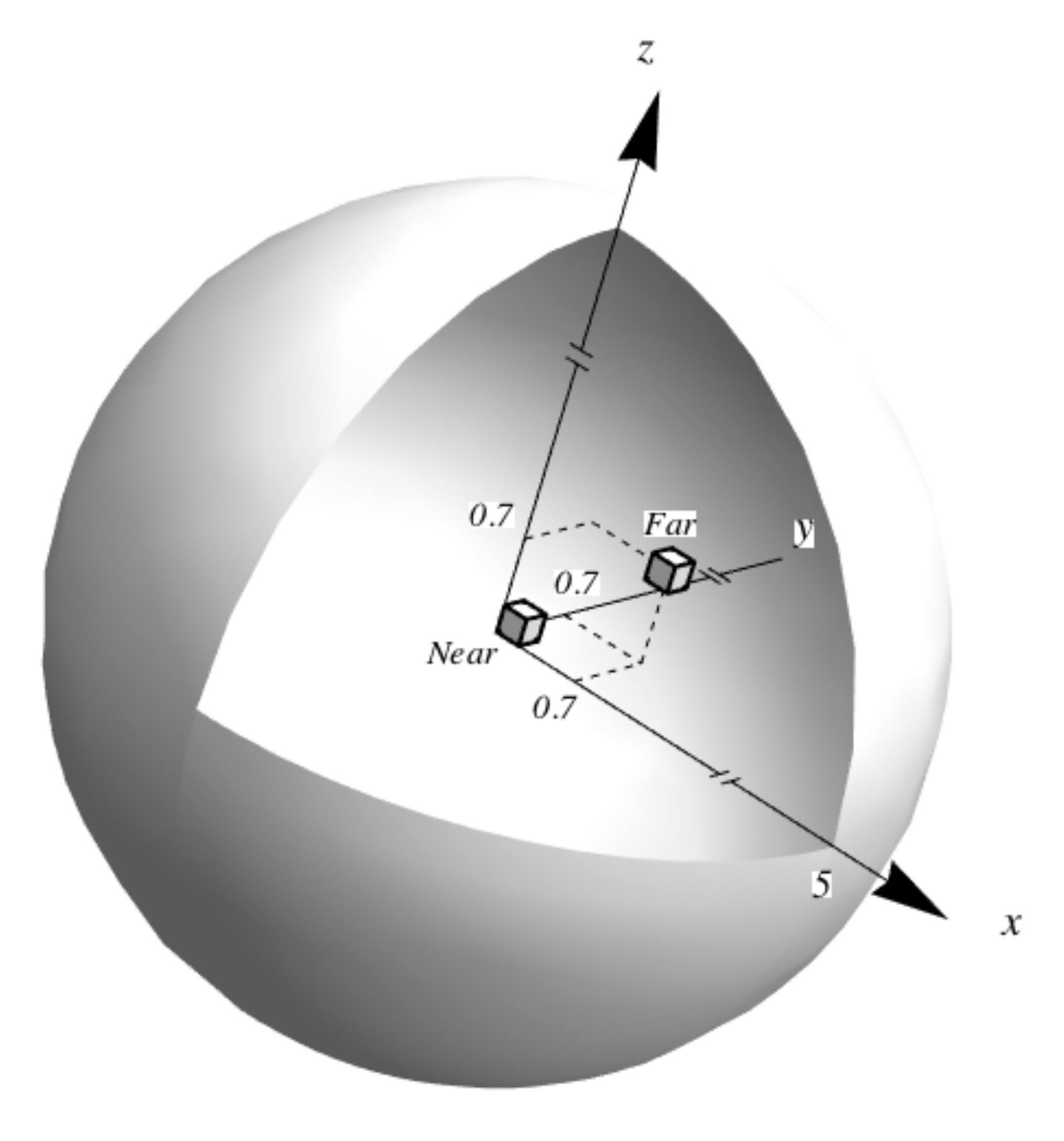}
\caption{The geometry of the "toy Monte Carlo". The source of transported particles is at the centre. The enclosing sphere has been scaled to make visible the active volumes for the track length observable used.}
\label{Fig:over1}
\end{figure}

We illustrate here an application of the method for uncertainty quantification
elaborated in the previous sections.
The calculation is performed in a streamlined simulation environment, which is
identified in the following as ``toy Monte Carlo'': although simplified, this
computational scenario implements all the essential elements of the conceptual
framework of uncertainty quantification illustrated in Fig.~\ref{Fig:over1}.

This example of application serves also as an environment for heuristic
analysis, with the purpose of identifying methods of more general applicability.

\subsection{Toy Monte Carlo}
\label{toymc}

The toy Monte Carlo is a lightweight simulation system,
interfaced to the computational environment of Mathematica 
\cite{mathematica} for further elaboration of its results.
It is an instrument for verification and heuristic investigation of theoretical
features of Monte Carlo particle transport relevant to the problem of
uncertainty quantification; it is not intended to implement physics and
experimental modeling functionality for real-life simulation application
scenarios.

It consists of a random path generator, which is ruled by two
constant parameters, describing the relative probability of absorption
($\Sigma_A$) and scattering processes ($\Sigma_S$) a transported particle
may undergo.
The user is free to modify these parameters.
This simplified situation physically corresponds to the transport of neutral
particles through a uniform medium, with energy independent absorption and
scattering cross sections, the scattering occurring always in S-wave (i.e.
scattering angles in the center-of-mass frame are sampled from an uniform
distribution).

Within this streamlined simulation context the units in which lengths and
macroscopic cross sections are expressed are not relevant; therefore for
simplicity we use arbitrary units for these two quantities, with the
only constraint of being one the reciprocal of the other.

The experimental configuration consists of an isotropic primary
particle source located at the origin of a sphere of uniform material.
The radius of the sphere is 5 in arbitrary units.

In the application example considered here we focus on the track length as an
observable ($O_{\rm tl}$ in the following) of interest in the simulation;
nevertheless the issues discussed in the following are not specific to this
observable, rather they highlight general features of the process of estimating
the uncertainty associated with a result produced by the simulation.
The observable is scored in two sensitive volumes, consisting of a cube of side
0.2 in arbitrary units: one with center located in (0.1,0.1,0.1), and one
centered in (0.7,0.7,0.7).

The execution of the Monte Carlo simulation contributes to the result through the
statistical sampling process.
If the physical parameters $\Sigma_A$ and $\Sigma_S$ are affected by
uncertainties, we expect the simulation outcome to be sensitive to the effects
of parameter uncertainties mixed with algorithmic uncertainties.

We stress that this toy Monte Carlo is only used to verify and to exemplify
results that have been derived on a theoretical ground in section \ref{sec_theory}:
its use does not constitute a limitation to the validity of the theoretical
assessments established in this paper, rather it represents an auxiliary
tool to our investigation.

\subsection{Algorithmic uncertainties}
\label{sec_appl_algorithmic}

\begin{figure}[!tbp]
\centering
\includegraphics[width=3.5in]{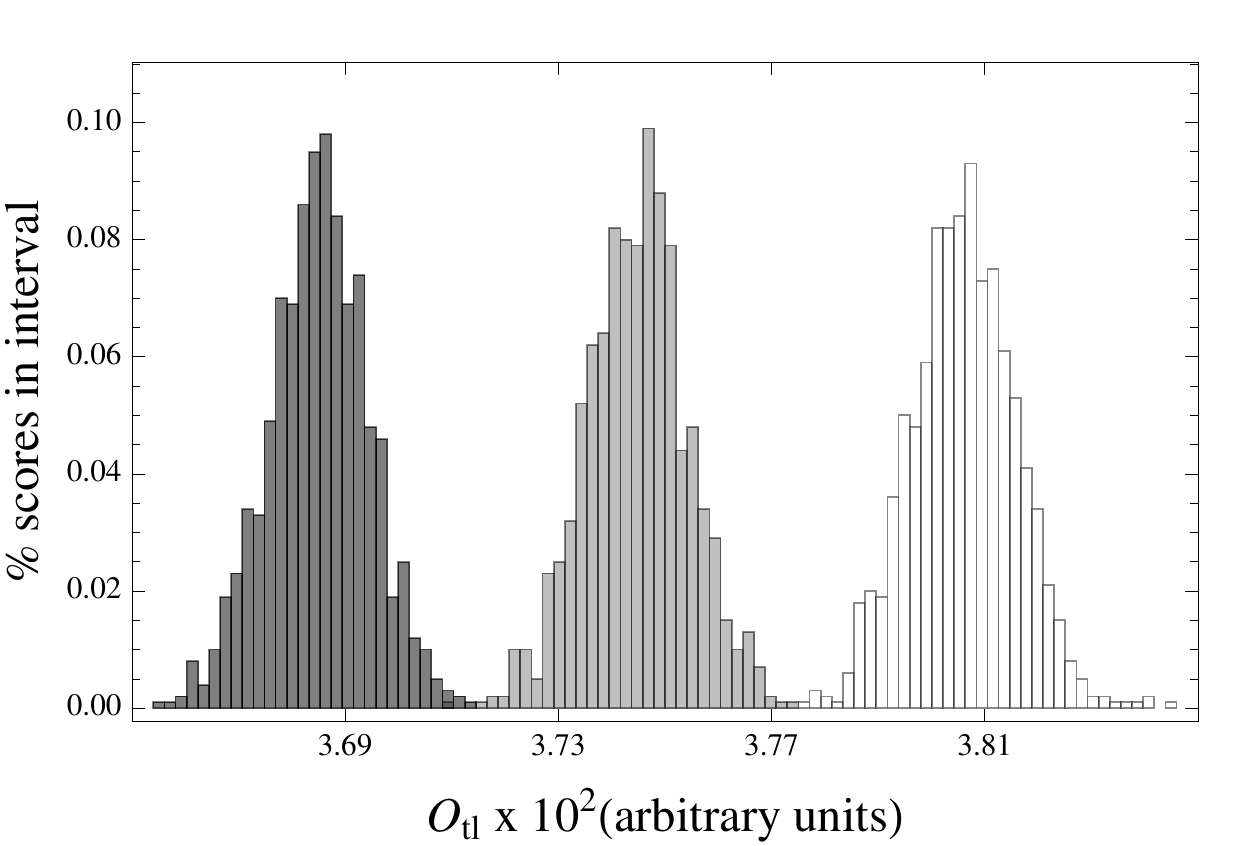}
\caption{Results for 1000 of Monte Carlo simulations for the observable $O_{\rm tl}$, each encompassing $10^6$
events, for an observable scored close to the primary particle source (see
text), produced with different values of the $\Sigma_S$ input physical
parameter: $\Sigma_S=1$ (white histogram), $\Sigma_S=1.1$ (grey
histogram) and $\Sigma_S=1.2$ (black histogram).}
\label{fig_1}
\end{figure}

\begin{figure}[!ltbp]
\centering
\includegraphics[width=3.5in]{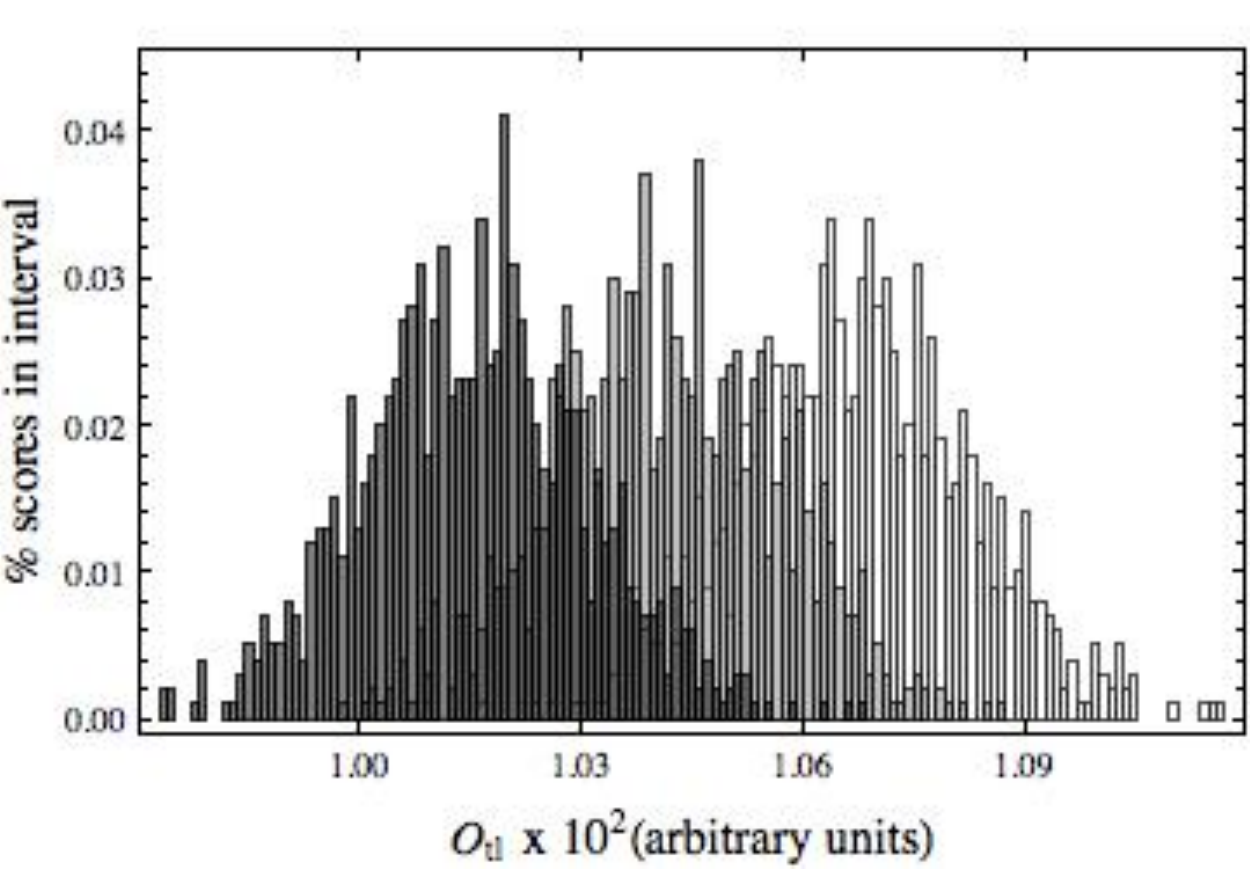}
\caption{
Results for 1000 of Monte Carlo simulations for the observable $O_{\rm tl}$, each encompassing $10^6$
events, for an observable scored far away
from the primary particle source (see
text), produced with different values of the $\Sigma_S$ input physical
parameter: $\Sigma_S=1$ (white histogram), $\Sigma_S=1.1$ (grey
histogram) and $\Sigma_S=1.2$ (black histogram).
}
\label{fig_1a}
\end{figure}

\begin{figure*}[ht]
 \centering
\mbox{
 \subfigure{
  \includegraphics[width=8.5cm]{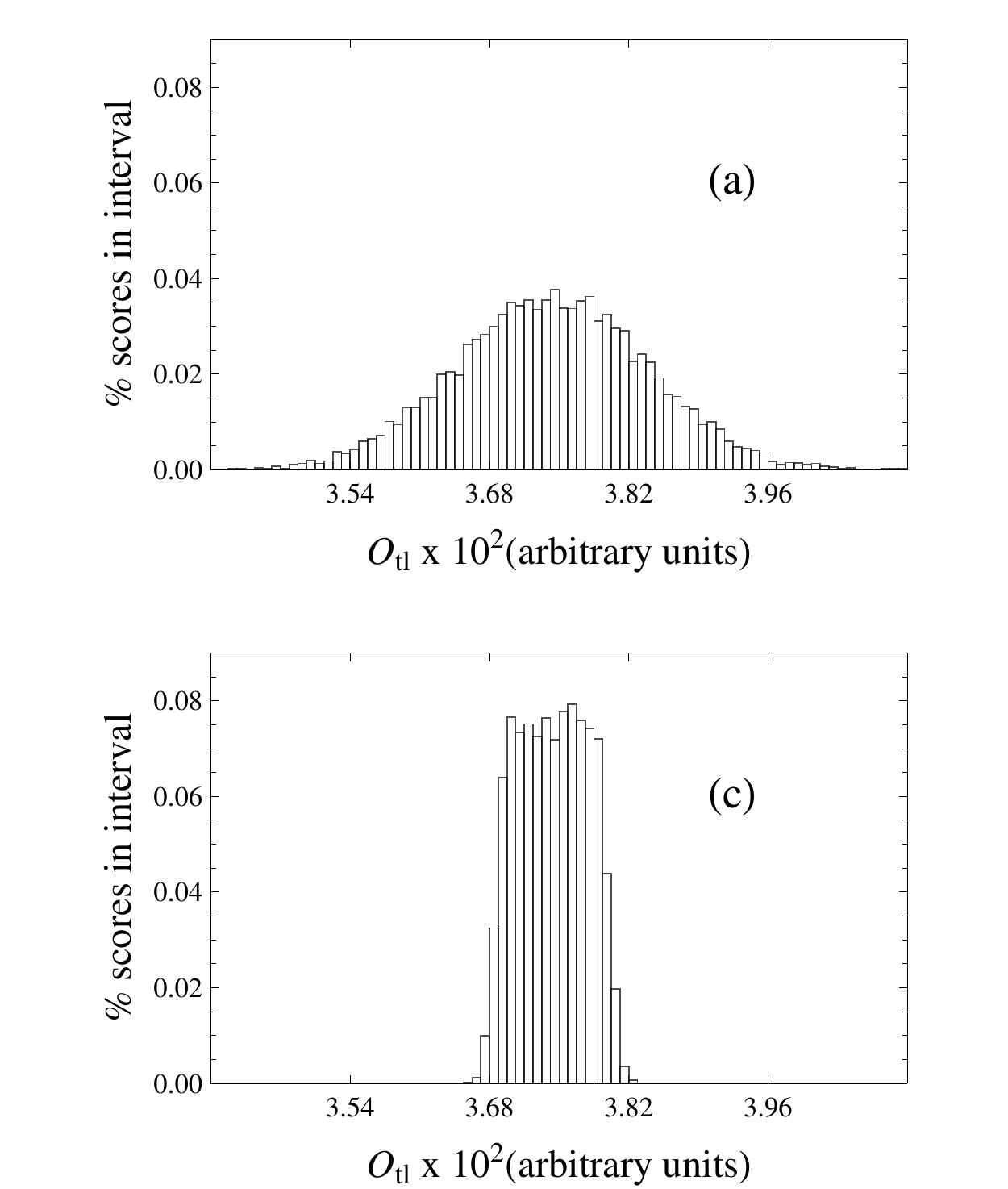}
   \label{fig_2papa}
   }
\quad
 \subfigure{
  \includegraphics[width=8.5cm]{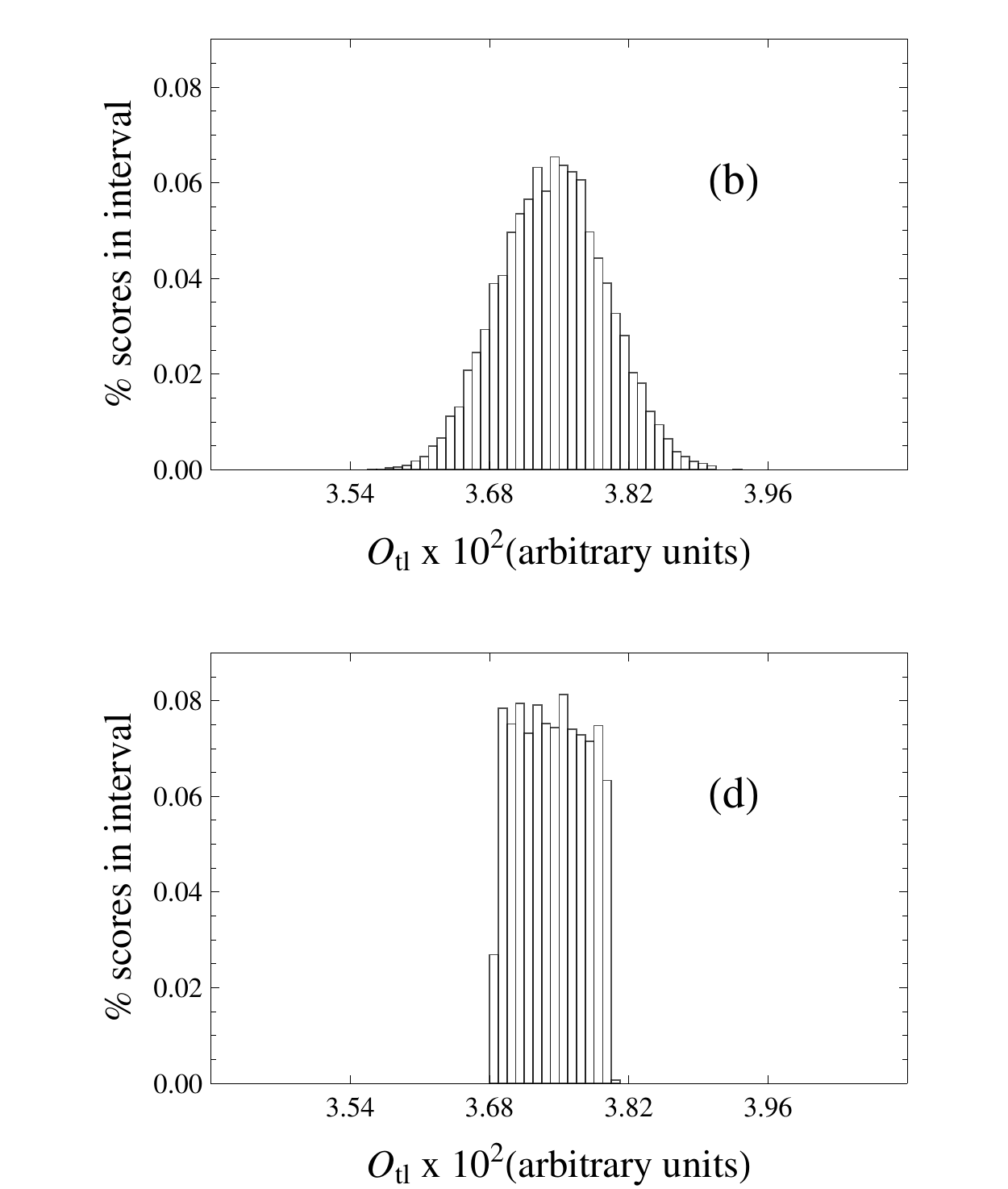}
   \label{fig_2papb}
   }}
\caption{Empirical distributions  of the track length observable $O_{\rm tl}$ scored close to the primary particle
source obtained by $10^5$ MC simulations.
Each single run has been produced with a random value for $\Sigma_S$ in the
interval $1.\le\Sigma_S\le 1.2$; each simulation consists of (a) $10^5$ events,
(b) $5\,10^5$ events, (c) $10^6$ events, (d) $10^8$ events.}
 \label{fig_22a}
\end{figure*}

The events generated in the course of a Monte Carlo simulation contribute to
determine observables of experimental interest.
Each particle trajectory encompassed in an event is the result of a sequence of
random interactions sampled from some statistics; the observable score from
single events itself is a stochastic variable with some unknown distribution.
The outcome of the Monte Carlo simulation is the mean of the contributions
from single events.
Nevertheless, we cannot directly determine the mean, the variance, the standard
deviation or any other properties of the distribution of the observable exactly,
because we do not know the distribution itself: we can only estimate these
quantities from the statistical sample.
As a consequence, the observable mean estimate itself is a stochastic variable.

As a first example, we illustrate the effect of algorithmic uncertainties, i.e. 
of uncertainties purely due to the Monte Carlo sampling process.
We consider three physical scenarios, where $\Sigma_S$ assumes the values
1.0,~1.1 and 1.2 respectively (in arbitrary units), while $\Sigma_A$ is fixed and
has value~0.1 (in arbitrary units).
In this use case the physical parameters $\Sigma_S$ and $\Sigma_A$
are assumed to be exempt from uncertainties.
For each physical scenario we execute 1000 Monte Carlo runs, each one encompassing
$10^6$ events corresponding to the transport of a single primary particle.
The resulting distributions of the observable, corresponding to the three
different values of $\Sigma_S$ assumed in the simulation, are plotted
in Fig.~\ref{fig_1} and Fig.~\ref{fig_1a} for the sensitive volume at closer and
larger distance from the source mentioned in section \ref{toymc}, respectively.
For graphical reasons here and in the following we make use of the unusual notation
$O_{\rm tl}\times 10^m$, even if the observable scored is in arbitrary units: this notation
has the purpose to indicate that the same units (even if arbitrary) have been used in all the plots; 
the indicated multiplier means that in Fig.~\ref{fig_1} the $x$-scale starts from $0.0369$ while
in Fig.~\ref{fig_1a} starts from $0.001$.

If the number of generated events is sufficiently large, the Central Limit
Theorem ensures that, whatever is the unknown observable distribution, the
distribution of the observable mean estimates produced by running many Monte
Carlo simulations is approximately gaussian, with standard deviation
\begin{equation}
\sigma_{<O_{\rm tl}>}=\sigma_{O_{\rm tl}}/\sqrt{N}
\end{equation}

The two experimental configurations, characterized by different distances
between the source and the detector, give rise to different statistical errors,
 because a larger number of events contribute to the observable score, 
if the sensitive volume is closer to the source.

\subsection{Effect of cross section uncertainties}
\label{sec_appl_cross}

\begin{figure*}[!t]
 \centering
\mbox{
 \subfigure{
  \includegraphics[width=9cm]{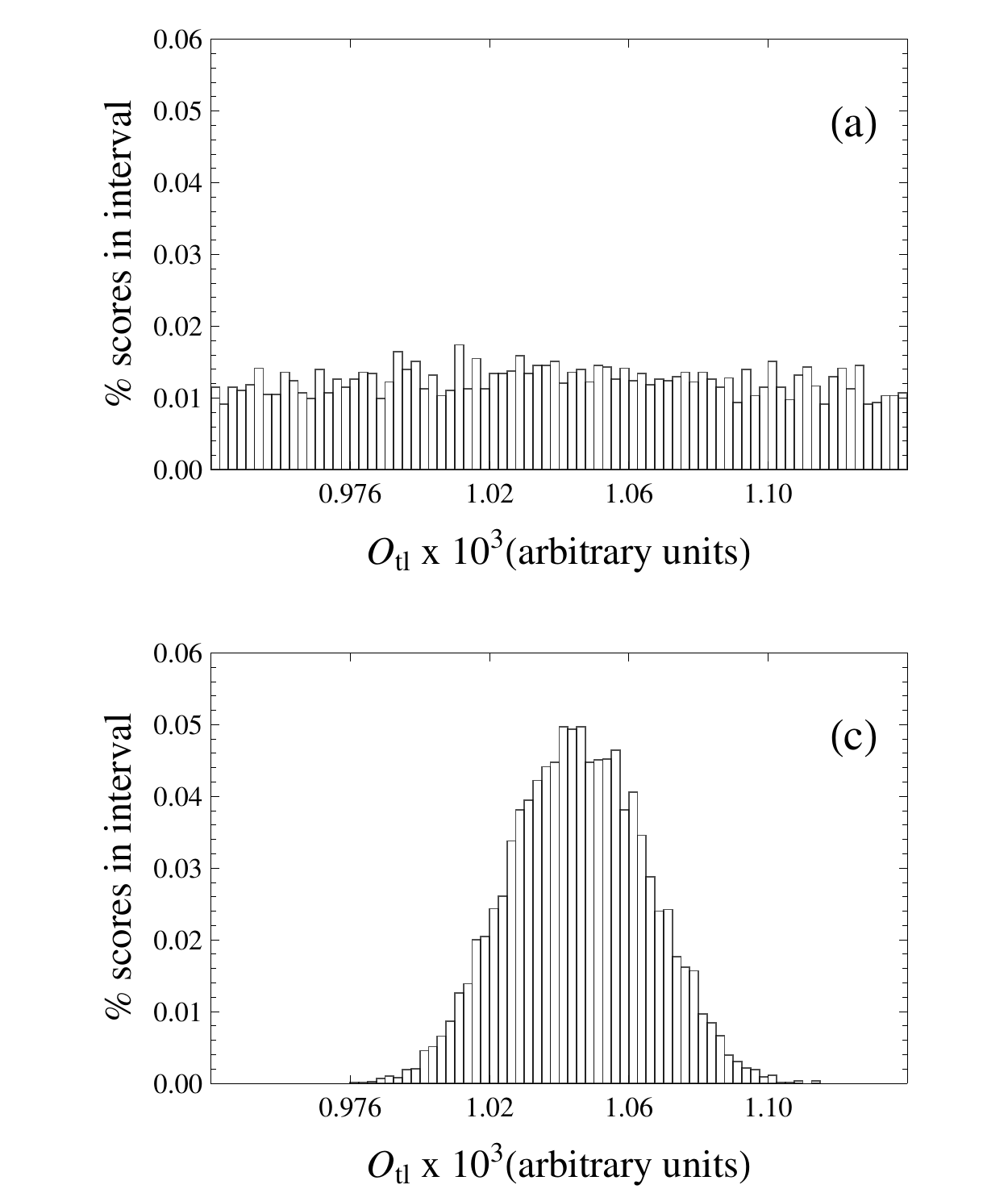}
   \label{fig_3papa}
   }
\quad
 \subfigure{
  \includegraphics[width=9cm]{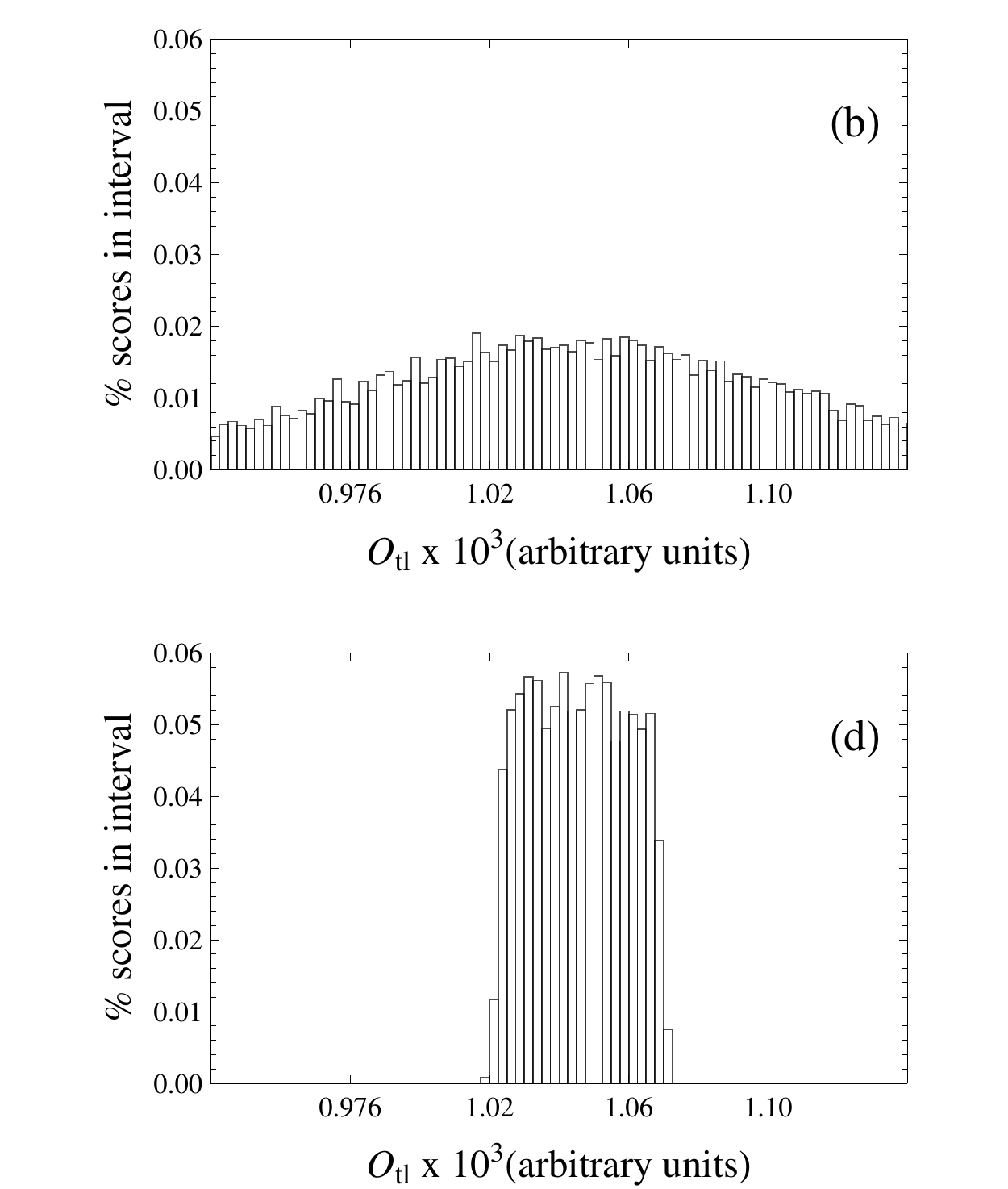}
   \label{fig_3papb}
   }}
\caption{
Empirical distributions  of the track length observable $O_{\rm tl}$ scored from the source (see
text) obtained by $10^5$ MC simulations.
Each single run has been produced with a random value for $\Sigma_S$ in the
interval $1.\le\Sigma_S\le 1.2$; each simulation consists of (a) $10^5$ events,
(b) $5\,10^5$ events, (c) $10^6$ events, (d) $10^8$ events.
}
 \label{fig_33}
\end{figure*}

Within the context of the toy Monte Carlo, we consider a scenario where one of
the input parameters is affected by uncertainties: its value may vary over some
interval with some probability distribution.
For simplicity we assume it to be a uniform distribution, which is corresponds
to the typical scenario of an epistemic uncertainty.

In this application example $\Sigma_S$ is sampled from a flat distribution in
the interval $1.\le\Sigma_S\le 1.2$.
We perform $10^4$ Monte Carlo simulation runs, each one with a different value of
$\Sigma_S$; each simulation encompasses a predefined number $N$ of events.
As described in section \ref{toymc}, the track length observable is scored in 
two volumes at different distances from the primary particle source.

Fig.~\ref{fig_22a} shows the distribution of the estimated observable means
close to the source resulting from the simulation; the four histograms
correspond to different number $N$ of events in each simulation run.
For a relatively small number of generated events, illustrated in the top row of
histograms in Fig.~\ref{fig_22a}, the distribution of the observable mean
estimates resembles a gaussian distribution.
By increasing the number of generated events from Fig.~\ref{fig_22a}(a) to
Fig.~\ref{fig_22a}(b), the width of the distribution decreases (roughly with
$\sqrt N$).
For larger values of $N$ the width stabilizes to a constant value and the
histogram no longer resembles a gaussian distribution: eventually it appears
consistent with a flat distribution.

Fig.~\ref{fig_33}, corresponding to the observable scored at larger distance
from the primary particle source, exhibits the same qualitative behavior,
although the process of approaching a final stable distribution appears slower
in this case.
This is a consequence, as noted in section~\ref{sec_appl_algorithmic}, of the larger
statistical error associated with sensitive volumes far from the source.

This example shows qualitatively that, apart from some additional statistical noise,
the distribution of observables resulting from the Monte Carlo simulation
process retains a functional form that is related to the input distribution of
the uncertainties of the parameters on which they depend.
These empirical observations are consistent with the theoretical foundations
discussed in section \ref{sec_theory}.

\subsection{Probability distribution function of an observable}
\label{sec_appl_pdf}

Equation (\ref{eqn:fin}) indicates that the task of uncertainty quantification is 
reduced to the determination of the output PDF $G(x)$.
From this distribution one can extract all the necessary information to quantify the 
knowledge of the output physical variable $x$ produced by the simulation, such as
the desired confidence intervals.
In the following we show how one can determine the characteristics of the PDF
of the simulated observable with predefined accuracy in the simple application
scenario introduced in section \ref{sec_appl_cross}. To do this we examine the properties of
the distribution from which simulation results are actually sampled - eq. (\ref{eqn:start}) - that 
we indicate here as $G_{\rm emp}(x)$, to distinguish it from the exact result $G(x)$.

In the case of a flat distribution of the input parameter
$f\left(\Sigma_S\right)=1/(\Sigma_{S\,{\rm max}}-\Sigma_{S\,{\rm min}})$ over an
interval $\Sigma_{S\,{\rm min}}\le\Sigma_S\le\Sigma_{S\,{\rm max}}$,  
the integral in equation (\ref{eqn:start}) can be evaluated
analytically even for finite $N$, if we assume that $\sigma_{x_0}=\sigma$ is
independent from $\Sigma_S$ and that $\Sigma_S(x)$ is linear.
The first assumption is justified, if one considers results at the lowest order
in $1/\sqrt N$ \cite{Papo};
we recall that $\sigma_{x_0}$ is the standard deviation of the required
observable (not the one of its mean), as it results from the simulation.
The second assumption is discussed in detail in section~\ref{sec_appl_linear}.
Under these two conditions the empirical distribution coming from the Monte Carlo
simulation resulting from (\ref{eqn:start}) is:
\begin{eqnarray}
G_{\rm emp}(x)&\simeq&\frac{\displaystyle d\Sigma_S/dx}{\displaystyle 2(\Sigma_{S\,{\rm max}}-\Sigma_{S_{\rm min}})}
\left[{\rm erf}\left(\frac{\displaystyle\sqrt{N}(x-a)}{\sqrt{2}\sigma}\right)\right.\nonumber\\ && \left.
-{\rm erf}\left(\frac{\displaystyle\sqrt{N}(x-b)}{\sqrt{2}\sigma}\right)
\right]\nonumber
\end{eqnarray}
being $${\rm erf}(x)=\frac{\displaystyle 2}{\sqrt\pi}\int_{-\infty}^x\exp[-t^2]dt\,;$$
or
\begin{eqnarray}
G_{\rm emp}(x)&\simeq&\frac{\displaystyle 1}{\displaystyle 2(b-a)}
\left[{\rm erf}\left(\frac{\displaystyle\sqrt{N}(x-a)}{\sqrt{2}\sigma}\right)\right.\nonumber\\ && \left.
-{\rm erf}\left(\frac{\displaystyle\sqrt{N}(x-b)}{\sqrt{2}\sigma}\right)
\right]
\,
\label{eqn:emp}
\end{eqnarray}
In equation (\ref{eqn:emp}) $a,b$ are the extrema of the range of variability of
the observable mean, or $a=x_0(\Sigma_{S,\,{\rm min}})$ and
$b=x_0(\Sigma_{S,\,{\rm max}})$.

\begin{figure}[!tbp]
\centering
\includegraphics[width=3.5in]{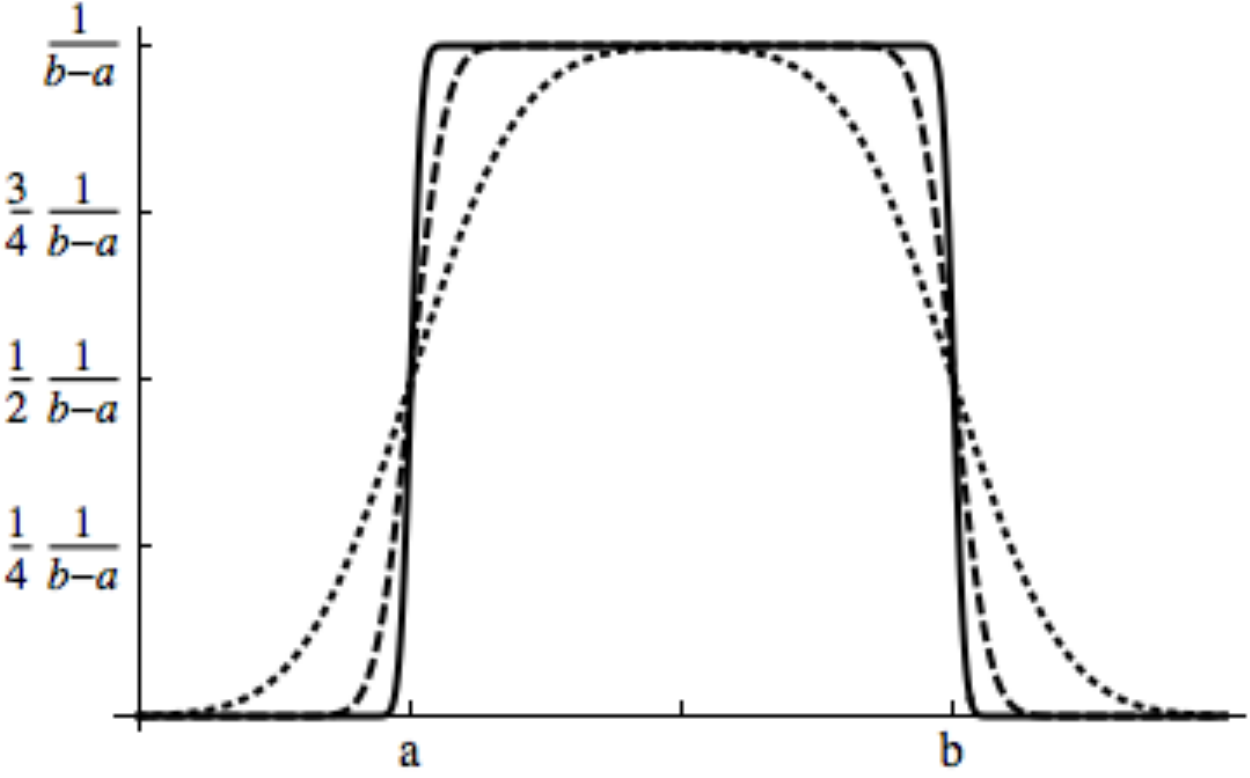}
\caption{The form of $G_{\rm emp}(x)$ for different values of $\sigma^2/N$: $10^{-3}$ dotted line, $10^{-4}$ dashed line and $10^{-5}$ solid line.}
\label{fig_errfun}
\end{figure}

The behavior of $G_{\rm emp}(x)$ is illlustrated in Fig.~\ref{fig_errfun} for
different values of $\sigma^2/N$: one can observe that the larger the value of
simulated events $N$, the more closely the curve resembles a flat distribution, which
reflects the flat distribution of the input parameter.
Fig.~\ref{fig_errfun} shows that the behavior of the empirical
PDF is directly ruled by the sole parameter $\sigma^2/N$.

The behavior of the analytically calculated $G_{\rm emp}(x)$
appears consistent with the outcome observed in the toy Monte Carlo.
The key point to note is that $G_{\rm emp}(x)$ from equation~ (\ref{eqn:emp}) is
the distribution from which observable means are actually sampled:
by running many Monte Carlo simulations one can obtain a statistical sample from it,
as is illustrated in Figs.~\ref{fig_22a} and \ref{fig_33}.
The histograms in these figures, which derive from the execution of a finite
number of Monte Carlo simulations, are approximations of analytical curves
as shown in Fig.~\ref{fig_errfun}.
The form they assume derives from the interplay between two different
statistical errors: the one coming from each Monte Carlo run ($\sigma^2/N$) and
the one coming from the finiteness of the number of Monte Carlo runs, as noted earlier in Section\,\ref{sect:verify}.

\begin{figure*}[thb]
\centering
\includegraphics[width=16cm]{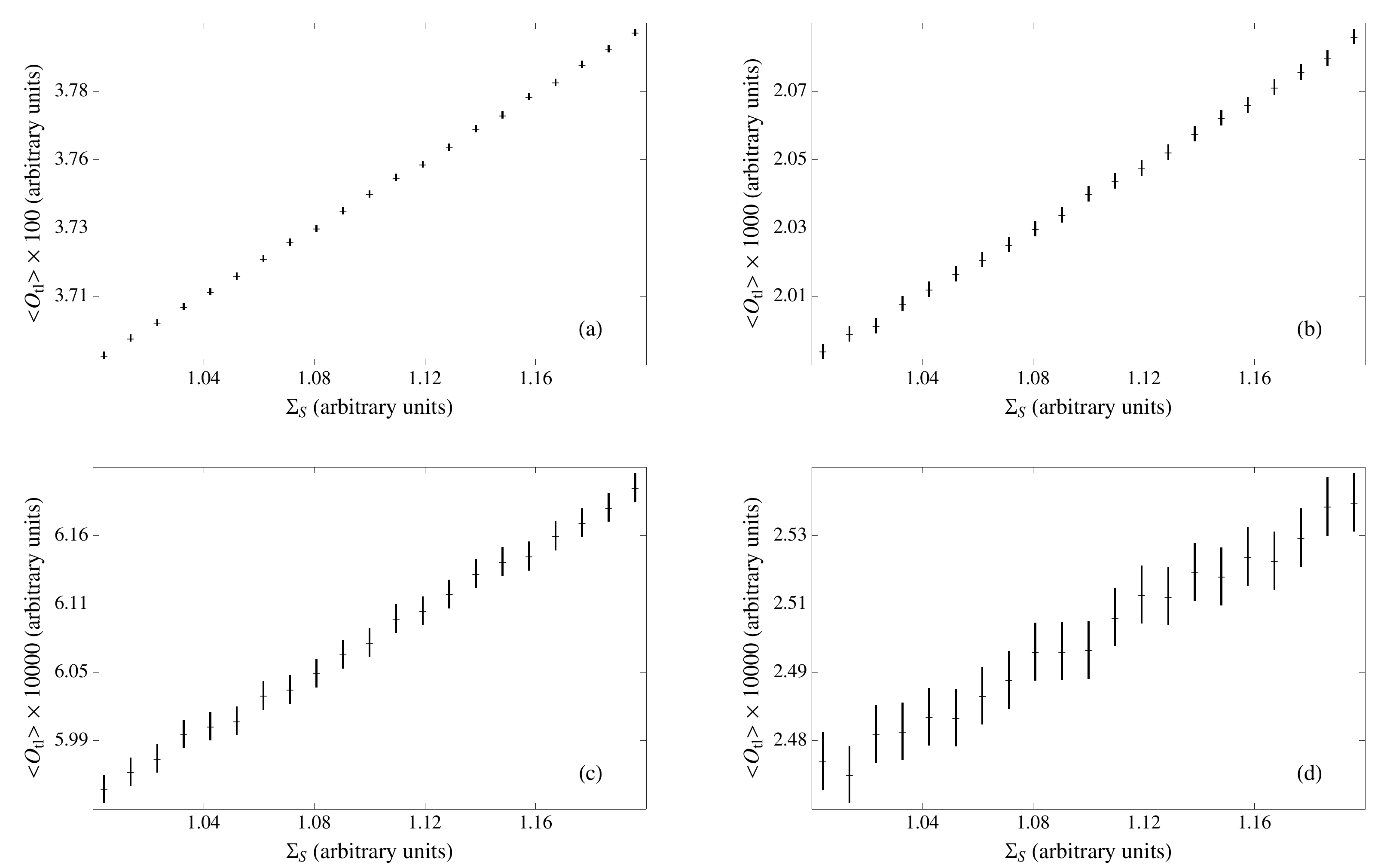}
\caption{The test on the form of the function $\Sigma_S(x)$: we plot for 21 equidistant and fixed values of $\Sigma_S$ the outcomes of a MC run, each  encompassing $10^8$ events, together with the corresponding statistical errors.
The four plots correspond to different distance of the detector from the primary
particle source: the center of the detector is placed at (a): 0.1, (b): 0.5 , (c): 0.9
and (d) 1.3 (in arbitrary units).
In the plot we show also the best linear fit to the data (see text).}
\label{fig_lin}
\end{figure*}

\subsection{Epistemic uncertainty of an observable}
\label{sec_appl_ab}

In some experimental scenarios knowledge of the range of variability of the
simulation outcome, rather than complete quantification of the uncertainty of
the observable mean, would be sufficient.
This range of variability can be considered as the epistemic uncertainty of the
observable produced by the simulation; for such a concept to be meaningful in a
context of uncertainty quantification, one should be able to estimate a priori
the statistical indetermination by which it is affected.
In the application scenario considered here this requirement corresponds to
knowing the statistical indetermination of extrema of the interval $[a,b]$.

For this purpose we observe that for any fixed value of $z$ and
$0<\varepsilon<1$, a value $\delta>0$ exists, such that
\begin{equation*}
\left|{\rm erf}(\sqrt N (x-z)/(\sqrt 2 \sigma)\right|<\varepsilon\;\iff\;\left|x-z\right|<\delta\,.
\end{equation*}
In particular, $\delta= \sqrt 2\sigma{\rm erf}^{-1}(\varepsilon)/\sqrt N$. 
In this interval the error function differs from $\pm 1$ more 
than $1-\varepsilon$.
The value of $N$, i.e. the number of generated events, fixes the statistical
indetermination on the evaluation of the true epistemic uncertainty $[a,b]$ 
of the simulated observable.
One can choose $N$ such that the intervals in $x$ over which both error
functions in (\ref{eqn:emp}) differ from $\pm 1$ by a predetermined amount are
as small as requested: to reduce $\delta$ one must increase $N$.
In practical terms, this means that the interval of variability of the
observable can be determined with any predefined statistical accuracy by two
Monte Carlo runs, corresponding to the values $\Sigma_S=\Sigma_{S,\,{\rm min}}$ 
and $\Sigma_S=\Sigma_{S,\,{\rm max}}$ of the input parameter, encompassing
an adequate number of events $N$.

From the previous discussion one can evince that for $G_{emp}(x)$ to represent 
an adequate approximation of a flat distribution the condition 
\begin{equation}
\sigma_{<O>}=\sigma/\sqrt N\ll b-a\,
\label{eqn:cond}
\end{equation}
should be satisfied.
This condition relates the scale of
$\sigma_{<O>}=\sigma/\sqrt N$ (that is the accuracy of a single Monte Carlo run)
and the range of variability of the observable mean.

\subsection{Approximation of the parametric dependence of an observable from an input unknown}
\label{sec_appl_linear}

In the previous discussion we have assumed a linear relationship $x_0(\Sigma_S)$
between the peak position of the observable means distribution and the input
parameter.
This assumption can be verified by running a few Monte Carlo simulations for
some values of the input parameters within its range of variability.

The results of performing this procedure in the context of our toy Monte Carlo
are illustrated in Fig.~\ref{fig_lin}.
Each plot shows the outcome of the simulation for 21 equidistant fixed values of
$\Sigma_S$ in the range of variability $1.0\le\Sigma_S\le 1.2$ together with
their $3\sigma$ confidence intervals. 
We display $3\sigma$ intervals instead of conventional $\sigma$ intervals for
better visibility in the plots.
The four plots correspond to different positions of the detector with respect to
the primary particle source, with center located at 0.1, 0.5, 0.9 and 1.3
arbitrary units from the source.
The number of events generated in each simulation is $N=10^8$. 
The result of a linear fit to the simulated data is superimposed to the plots.
A linear relation $x_0(\Sigma_S)$  qualitatively appears a justified approximation, although
the statistical degradation of the quality of a linear fit is clearly visible,
when the observable is scored at larger distance from the source.
It is quantitatively supported by the p-values of the linear fit: 0.995, 0.993,
0.997 and 0.985 respectively, for increasing distance from the source.
At increasing distances of the detector from the source a larger number of
events can be generated, if necessary, to reduce the size of the confidence
intervals for the parameters of the fit: in practical cases a detailed
statistical analysis is recommended, since the knowledge of the expected PDF
$G(x)$ is affected by the uncertainties in the best-fit parameters for the
relation $x_0(\Sigma_S)$.

This method can be used more generally to investigate possible approximations
of the $x_0(\Sigma_S)$ functional relationship.
The resulting fitted function $x_0(\Sigma)$ can be inserted directly in equation
(\ref{eqn:fin}) to obtain the output PDF.

\section{Method for estimating output uncertainties}
\label{sec_methods}

The theoretical findings of section \ref{sec_theory} and the heuristic
investigation of section \ref{sec_application} in a simple application
environment support the definition of a path for the calculation of the
uncertainties of simulation observables in a generic scenario.
We summarize it here from the practical perspective of a Monte Carlo simulation
user.
This procedure assumes that the uncertainties of input parameters are known.

Equation (\ref{eqn:fin}) states that the form of the output PDF is assured by
theory.
As a consequence, the process of uncertainty quantification is reduced to
determining the parameters required for its calculation: this problem in turn
consists of determining the unknown function $x_0(\Sigma)$, which represents the
response of the observable means to the variation of the input parameter, or its
inverse $\Sigma(x_0)$.

From a Monte Carlo user perspective, the process involves the investigation of
two properties of the output.

The first concerns the extrema of the output variability interval for the
observable means, that is the maximum output uncertainty.
These extrema can be found with an arbitrary predefined precision
with two Monte Carlo runs, the first using as input $\Sigma=\Sigma_{\rm
min}$, the second using as input $\Sigma=\Sigma_{\rm max}$.
We stress that this is not a full quantification of the uncertainty, if nothing
is known about the relation $x_0(\Sigma)$;
these two values, together with the statistical uncertainties on their
determination, define an interval of variability for the output, not its
probability distribution inside this interval.

A remark is here necessary, regarding the possibility of an input PDF not
supported on a bounded interval - this is the case for instance of a normal distribution
which is supported on $(-\infty,+\infty)$:
we suggest in this case to first identify
an interval of interest for the input, for instance associated with a given
confidence interval for the input, and to proceed on with the analysis.
In this case one will not obtain the output PDF, but the output PDF conditioned to
the probability of the input to be in this interval.

Approximate knowledge of the function $x_0(\Sigma)$ can be achieved
through statistical investigation, by performing a number of Monte Carlo
simulations, involving different values of the input over its variability
interval (or over the chosen confidence interval for the input unknown),
 to devise a suitable functional approximation.
How to select these simulation configurations
is a problem-dependent task; in the absence of any specific indications, a
proper starting choice could be a set of equidistant points, but an adaptive
procedure could be useful or necessary, especially if $x_0(\Sigma)$ is not
linear or if the input PDF is not flat.
This process profits from knowing a priori that condition (\ref{eqn:cond})
must be fulfilled, as was elucidated in the discussion of the application
example in section \ref{sec_appl_ab}.

From this procedure one can estimate quantitatively confidence
intervals for the value of the output observable $x$.
The computational resources needed for this investigation depend on the required
accuracy of the estimate, but they are significantly lower than the
computational investment needed to obtain directly an empirical approximation to
$G(x)$: the above outlined procedure involves a limited number of simulation
runs, while direct evaluation of the uncertainty of an observable according to
the procedure depicted in Fig.~\ref{Fig:over} requires executing a larger number 
$N_{MC}$ of simulations to produce a distribution of observable means with adequate
statistical precision. 
To give an example, we used $N_{MC}=10^5$ to obtain Fig.\ \ref{fig_22a}
and \ref{fig_33}.

\section{Conclusion}

We have shown that, under wide and verifiable conditions, the process of Monte
Carlo simulation transfers the input probability distribution function (PDF) of
the physical data on which the simulation depends into a predictable PDF
of the observable means, which are the outcome of the simulation.

Often, due to the narrowness of the variability interval involved, the same
functional form of the PDF may apply both to input and output uncertainties, by
means of a simple linear mapping, whose parameters can be deduced with
predetermined accuracy by running a small number of Monte Carlo simulations.

The process of uncertainty quantification is intertwined with the mathematical
method used to solve the problem that relates the input to the output: in Monte
Carlo particle transport uncertainty quantification is blurred by the stochastic
process of sampling, which is involved in transferring input uncertainties into
the output.

The procedure for uncertainty quantification here presented applies easily when
one, or a small number of, physical parameters are involved, and if they can be
independently analyzed: in experimental practice this scenario applies to
simulations where a few physics features play a dominant role in determining the
key observables subject to investigation.
The present approach is hardly practicable when many input physical data
can vary simultaneously.
Methods to extend it to a multidimensional case, also taking into account that
some variations may not be considered independent, are currently being developed
and will be documented in forthcoming publications: the present paper provides
the essential theoretical reference for extensions to more complex physical
scenarios and mathematical calculations.


\section*{Acknowledgment}
The authors thank the INFN Computing and Networking Committee (CCR)
for supporting this research.

The CERN Library, in particular Tullio Basaglia, has provided helpful assistance
and reference material for this study.

\appendices

\end{document}